\documentclass[twocolumn]{aastex631}

\usepackage{verbatim}

\newcommand{\hmsun}{\ensuremath{h^{-1}\,M_\odot}}
\newcommand{\hmpc}{\ensuremath{h^{-1}\,\mathrm{Mpc}}}

\newcommand{\delpc}{\ensuremath{\delta_F^{\mathrm{pc}}}}
\newcommand{\delw}{\ensuremath{\delta_F^{\mathrm{w}}}}
\newcommand{\kfloor}{\ensuremath{K_\mathrm{floor}}}

\submitjournal{ApJL}

\begin{document}

\title{Observational Evidence for Large-Scale Gas Heating  \\ in a Galaxy Protocluster at $z=2.30$}

\author[0000-0002-8505-9815]{Chenze Dong}
\affiliation{
Kavli Institute for the Physics and Mathematics of the Universe (WPI), UTIAS, \\
The University of Tokyo, Kashiwa, Chiba 277-8583, Japan
}

\correspondingauthor{Chenze Dong}
\email{chenze.dong@ipmu.jp}

\author[0000-0001-9299-5719]{Khee-Gan Lee}
\affiliation{
Kavli Institute for the Physics and Mathematics of the Universe (WPI), UTIAS, \\
The University of Tokyo, Kashiwa, Chiba 277-8583, Japan
}

\author[0000-0002-5934-9018]{Metin Ata}
\affiliation{
The Oskar Klein Centre, Department of Physics, Stockholm University,  \\ AlbaNova University Centre, SE 106 91 Stockholm, Sweden
}
\affiliation{
Kavli Institute for the Physics and Mathematics of the Universe (WPI), UTIAS, \\
The University of Tokyo, Kashiwa, Chiba 277-8583, Japan
}

\author[0000-0001-7832-5372]{Benjamin Horowitz}
\affil{Lawrence Berkeley National Laboratory, 1 Cyclotron Road, Berkeley, CA 94720, USA}
\affiliation{Department of Astrophysical Sciences, Princeton University, Princeton, NJ 08544, USA}

\author[0000-0002-8857-2905]{Rieko Momose}
\affiliation{
Observatories of the Carnegie Institution for Science, \\ 
813 Santa Barbara Street, Pasadena, CA 91101, USA
}
\affiliation{
Department of Astronomy, School of Science, \\ 
The University of Tokyo, 7-3-1 Hongo, Bunkyo-ku, Tokyo, 113-0033, Japan
}



\begin{abstract}
We report a $z=2.30$ galaxy protocluster (COSTCO-I) in the COSMOS field, where the Lyman-$\alpha$ forest as
seen in the CLAMATO IGM tomography survey does not show significant absorption. This departs from the transmission-density relationship (often dubbed the fluctuating Gunn-Peterson approximation; FGPA) usually expected
to hold at this epoch, which would lead one to predict strong Ly$\alpha$ absorption at the overdensity. 
For comparison, we generate mock Lyman-$\alpha$ forest  maps by applying FGPA to constrained simulations of the COSMOS density field, 
and create mocks that incorporate the effects of finite sightline sampling, pixel noise, and Wiener filtering. 
 Averaged over $r=15\,\hmpc$ around the protocluster, the observed Lyman-$\alpha$ forest is consistently more transparent in the real data than in the mocks, indicating a rejection of the null hypothesis that the gas in COSTCO-I follows FGPA ($p=0.0026$, or $2.79 \sigma$ significance).
%Other protoclusters in the volume generally show agreement between the mocks and data.
It suggests that the large-scale gas associated with COSTCO-I is being heated above the expectations of FGPA, which might be due to either
large-scale AGN jet feedback or early gravitational shock heating. 
%% This is the first known large-scale region of the IGM that is transitioning from the optically-thin photoionized regime at Cosmic Noon, to eventually collapse into the intra-cluster medium by $z=0$. 
COSTCO-I is the first known large-scale region of the IGM that is observed to be transitioning from the optically-thin photoionized regime at Cosmic Noon, to eventually coalesce into an intra-cluster medium (ICM) by $z=0$. 
Future observations of similar structures will shed light on the growth of the ICM and allow constraints on AGN feedback mechanisms.
\end{abstract}



\section{Introduction} \label{sec:introduction}
After the end of hydrogen reionization by $z\sim 6$, the vast majority of hydrogen in the universe is ionized and heated.
As photon heating is no longer effective in the optically-thin ionized intergalactic medium (IGM), 
the competition between photon heating and adiabatic cooling gradually erases the thermal history of reionization \citep{hui:2003, trac:2008}. By the ``Cosmic Noon" epoch of $z\sim 2-4$, this is expected to result in a universal power-law temperature-density relation for the IGM:
\begin{equation}\label{eq:tdr}
    T \propto (1 +\delta_m)^{\gamma - 1},
\end{equation}
where $T$ is the temperature of IGM, and $\delta_m = \rho / \bar{\rho} - 1$ is the underlying matter overdensity traced by the IGM. $\gamma$ is the power-law index of the temperature-density relation, which is expected to be $\gamma \approx 1.5$ from theoretical expectations \citep{hui:1997a}; 
at $z\sim 2-3$ this has largely been confirmed by Ly$\alpha$ forest observations \citep{lee:2015,hiss:2018,rorai:2018}.

The residual neutral hydrogen in the optically-thin, photoionized IGM is detectable as Lyman-alpha (Ly$\alpha$) forest absorption\footnote{In this paper, ``Ly$\alpha$'' absorption refers implicitly to optically-thin forest absorption; we will not discuss optically-thick absorbers.}, 
which is often approximated by an analytical relation between matter overdensity and Ly$\alpha$ optical depth $\tau$
\begin{equation}\label{FGPA}
    \tau \propto \frac{T^{-0.7}}{\Gamma_\mathrm{uv}}(1 + \delta_m)^2 \propto (1 + \delta_m)^{\beta},
\end{equation}
in which $\Gamma_\mathrm{uv}$ is the background ultraviolet (UV) photoionization rate, and the power-law index $\beta$ satisfies $\beta = 2 - 0.7 (\gamma - 1)$ after substituting in the temperature-density relationship of Equation~\ref{eq:tdr}.
While this power-law relation, often known as the fluctuating Gunn-Peterson approximation (FGPA), is not expected to be exact, comparisons with hydrodynamical simulations have found 
that it remains a useful heuristic \citep{peirani:2014,sorini:2016} and is useful over a wide range
of applications in the $z>2$ Ly$\alpha$ forest.   

At later times ($z<1.5$ or lookback times of $<9.5\,$Gyrs), FGPA is expected to gradually break down as large-scale shocks from non-linear gravitational collapse lead to collisional heating of the IGM. Simultaneously, feedback from galaxies and supermassive black holes are expected to deposit additional energy into the IGM, leading to a complex multi-phase IGM \citep{cen:2006} at $z\sim 0$ that still remains to be fully characterized \citep[e.g.,][]{shull:2012,de-graaff:2019}. 

Galaxy protoclusters --- progenitors of the massive galaxy clusters seen at late times -- are an interesting testbed for the evolution of cosmic gas, as they collapse earlier than less dense regions of the Universe, and host significant fractions ($>20\%$) of cosmic star-formation at high redshifts \citep{chiang:2017}. 
Early searches for $z\gtrsim 1$ protoclusters were  
dominated by searches around `signposts' such as radio galaxies or luminous quasars ---  leading to unrepresentative and incomplete protocluster samples. 
Over the past decade, however, `blind' searches in photometric or spectroscopic data have become more common (see \citealt{overzier:2016} for a review). 
Arguably the most sophisticated technique to-date is the application of density reconstructions and constrained simulations on $z\gtrsim 2$ galaxy redshift surveys covering representative cosmic volumes in the COSMOS field \citep{ata:2021,ata:2022}. 
This has allowed bespoke gravitational modelling of observed $2.0<z<2.5$ large-scale structures, hence enabling the discovery and characterization of protoclusters down to lower final masses ($M(z=0)\approx 4-6\times 10^{14}\,\hmsun$) than feasible with most other methods.

While the matter associated with low-redshift cluster halos occupy volumes of $\sim 1$Mpc$^3$, at $z\gtrsim 2$
their Lagrangian extent is of order $\gtrsim (10\,\mathrm{cMpc})^3$ \citep{chiang:2013}.
This is prior to the regime of fully non-linear collapse and gravitational shock-heating, therefore the Ly$\alpha$ absorption from $z\gtrsim 2$ protoclusters
is still expected to trace the density on $\sim$Mpc scales \citep{miller:2021}. 
This fact has motivated various searches of $z>2$ galaxy protoclusters through their Ly$\alpha$ forest absorption
\citep{stark:2015,lee:2016,cai:2016,ravoux:2020,qezlou:2022,newman:2022}.
%At the same time, the study of protoclusters promises important insights into the late-time properties of galaxy clusters, which is of great interest from both the astrophysical and cosmological points-of-view. 

In this Letter, we use the \citet{ata:2022} protocluster sample (and associated data products) in combination
with Ly$\alpha$ forest absorption data in the COSMOS field to show that the infalling gas associated with a $z=2.30$ galaxy protocluster appears to be heated beyond the expectations of the FGPA, over scales of multiple Mpc. 
%In Section \ref{sec:data}, we first introduce the COSTCO protocluster sample and constrained simulation suite and describe the basic properties of the specific protocluster we are analyzing. Then, we introduce the CLAMATO Ly$\alpha$ forest data used in our analysis.
%In Section \ref{sec:analysis}, we formulate our null hypothesis that the protocluster gas is obeying the FGPA by building mock data sets, which we show is not consistent with the observations.
%Finally, in Section \ref{sec:discussion}, we discuss our results in the context of previous observations and simulation work, 
%and briefly mention possible future directions. 
Throughout this paper, we adopt a cosmology
of $H_0 = 100\,h\,\mathrm{km\,s^{-1}\,Mpc^{-1}} = 70 \,\mathrm{km\,s^{-1}\,Mpc^{-1}}$, $\Omega_m = 0.315$, and $\Omega_\Lambda = 0.685$. To avoid confusion, we use $\hmpc$ as the unit of comoving distance, and pMpc when discussing physical scales.

\section{Data} \label{sec:data}
%In this section, we first discuss the $z=2.298$ galaxy protocluster that is the basis of our study, 
%followed by a brief description of the Ly$\alpha$ forest tomography data that probes its large-scale gas properties. 

\subsection{The COSTCO-I Galaxy Protocluster}
The possible presence of a protocluster at $z=2.30$ in the COSMOS field was
first noted as a compact overdensity of galaxies by \citet{lee:2016}, along with the unusually high Ly$\alpha$ transmission given the overdensity. However, no detailed analysis was carried out.

The protocluster was subsequently confirmed by \citet{ata:2022}. In this study, they applied the techniques of density reconstructions and constrained simulations to existing large-scale 
spectroscopic surveys that have targeted galaxies at the $z\sim 2-3$ epoch in the COSMOS field (e.g.\ zCOSMOS, VUDS, MOSFIRE; \citealt{lilly:2007}, \citealt{ le-fevre:2015}, \citealt{kriek:2015}). 
First, in \citet{ata:2021}, the \verb|COSMIC-BIRTH| hybrid Monte-Carlo density reconstruction algorithm \citep{kitaura:2021} was applied to estimate the underlying density field and corresponding initial density fluctuations (at $z=100$) that would eventually evolve to provide the best match for the $2.0<z<2.5$ spectroscopic galaxy distribution over the central $\sim 1$\,deg$^2$ of COSMOS.
This technique computes the Bayesian posterior probability of the possible initial conditions, thus sampling the uncertainties inherent in the observational data.
A subset of the initial condition realizations were used to seed numerical N-body `constrained' simulations  \citep[dubbed the `COSTCO' suite,][]{ata:2022} that were ran to $z=0$ to track the 
gravitational evolution of the density field traced by the observed galaxies. 
They then identified galaxy clusters with $M>2 \times 10^{14}\,\hmsun$ in the $z=0$ simulation snapshot, which were then matched to observed structures at $2.0<z<2.5$. 
This study confirmed several previously-known protoclusters in COSMOS 
such the ZFIRE protocluster at $z=2.095$ \citep{nanayakkara:2016} and the `Hyperion' proto-supercluster at $z\approx 2.45-2.50$ \citep{cucciati:2018}. In addition to these, 
a number of new protoclusters were also discovered.

\begin{figure*}[t!]
    \centering
    \begin{interactive}{js}{COSTCO_member_int.zip}
    \begin{centering}\includegraphics[width=0.62\textwidth,clip=true,trim=170 5 115 6]{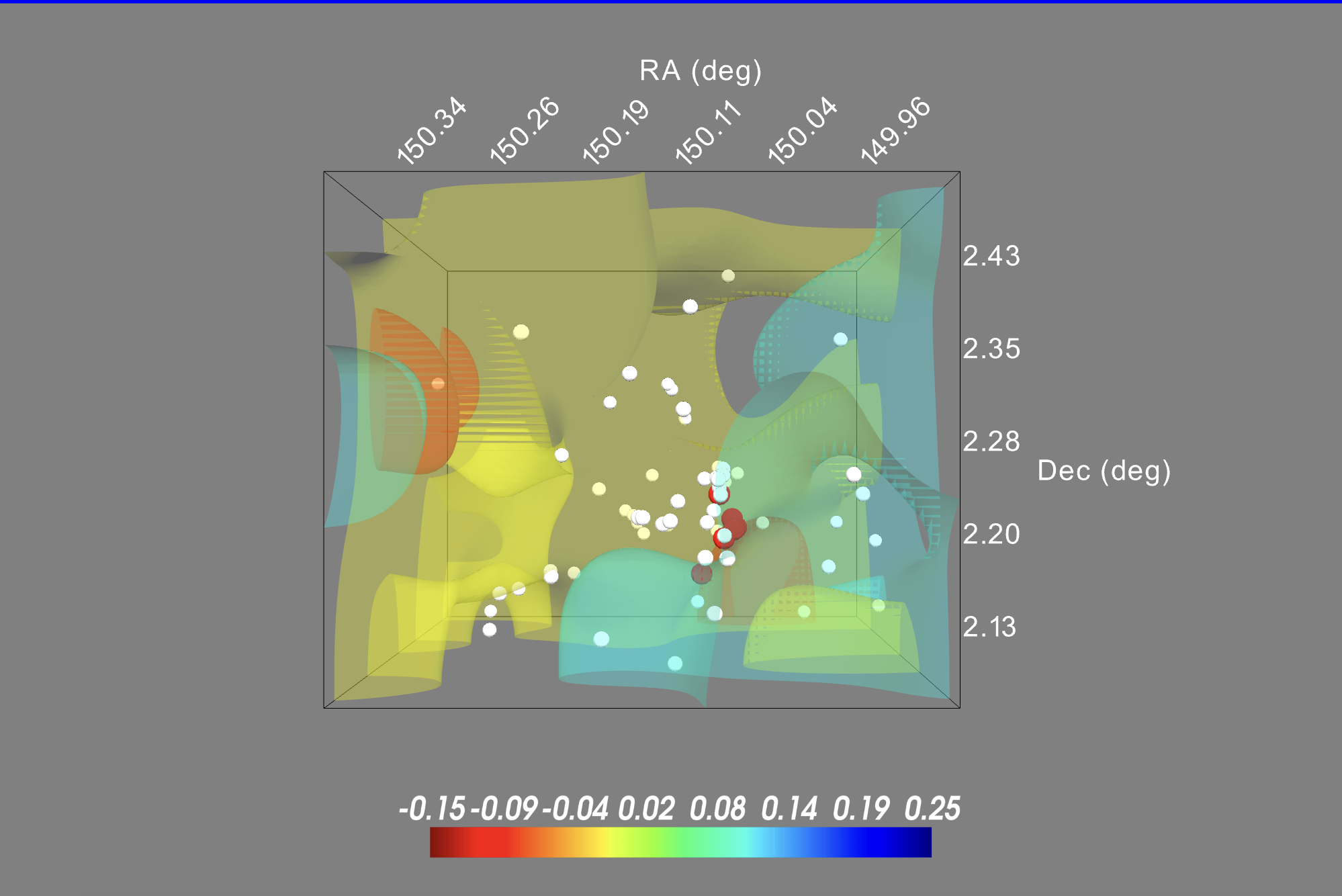}\end{centering}
    \end{interactive}
    \caption{
% TO AAS JOURNAL COPY EDITORS: For published version the first sentence in parentheses can be deleted or amended to point to published online figure.
%    (The interactive version is available online \href{https://member.ipmu.jp/chenze.dong/materials/COSTCO-I/}{at this URL}.) 
A preview of interactive plot (at this \href{https://member.ipmu.jp/chenze.dong/materials/COSTCO-I/}{url}) demonstrating the position of COSTCO-I members in the volume of CLAMATO. The red, yellow, cyan surfaces represent 
    Wiener-filtered Ly$\alpha$ flux of CLAMATO smoothed with 4 $\hmpc$ Gaussian filter, $\delw=-0.1, 0, 0.1$, respectively; the white dots are positions of galaxies in the catalog by \citet{Momose:2022}. We emphasize the core members of COSTCO-I with red spheres, and it is clear there is no Ly$\alpha$ excess (red contours) associated with the protocluster. The static preview shows the field in the transverse plane projected over redshift range $2.27<z<2.33$, but the online version is fully 3D with a redshift range of $2.18<z<2.41$. In the interactive figure, users can zoom in/out and/or rotate the figure about any axis. At the bottom of the interactive figure are 4 buttons. These allow the user to reset the view back to the default orientation, a view along Right Ascension, a view along Declination, and a view along redshift. \label{plot3d}}
\end{figure*}

COSTCO J100026.4+020940 (hereafter ``COSTCO-I"), located at RA = $150.110\degr  \pm  0.042\degr$, dec = $2.161\degr \pm 0.040\degr$ and $z=2.298\pm 0.007$, was one of the strongest detections  {(at 8.7$\sigma$ significance)} among these newly-discovered protoclusters. The final mass 
was estimated to be $M(z=0) = (4.6\pm 2.2) \times 10^{14}\,\hmsun $. 
While \citet{ata:2022} did not further examine COSTCO-I in detail, we
searched the spectroscopic redshift catalog compiled by \citet{Momose:2022} 
%% new method is depicted here
and identified member galaxies of COSTCO-I. 
Since the position of COSTCO-I varies among the realizations, 
we adopted the following approach to obtain a robust choice of core members: 
First, we usd the position of COSTCO-I reported by \citet{ata:2022} 
as an initial guess for the protocluster center, 
and selected galaxies within 
a $6\,\hmpc$ transverse radius 
and line-of-sight (LOS) velocity window of $ |\Delta v| < 600\,\mathrm{km\, s^{-1}}$. 
Then, we iteratively recalculated the protocluster center as the median position of the member galaxies
and repeated the member selection until convergence.
With this method, we found 7 galaxies in the vicinity of COSTCO-I.
Note that these galaxies are merely the putative collapsed core --- the full
Lagrangian extent that would eventually collapse into the $z=0$ cluster occupies
a larger extent than this.
Of these core galaxies, the most massive galaxy has a stellar mass of $M_* = 5.6 \times 10^{10}\,M_\odot$.
These galaxies are shown in Figure~\ref{plot3d},  {which is an interactive figure that can be viewed \href{https://member.ipmu.jp/chenze.dong/materials/COSTCO-I/}{online}
and also has been uploaded to \dataset[Zenodo]{https://doi.org/10.5281/zenodo.7559495}.}

After identifying the protocluster core, 
we estimated the group mass, $M_V$,  {in order to set the upper limit to the extent of the intra-group or intra-cluster medium that could already be present at $z=2.30$}.
We used the virial theorem approach outlined by
\citet{girardi:1998}: 
\begin{equation}
M_V = \frac{3\pi}{2} \frac{\sigma_p^2 R_p}{G};
\end{equation}
$G$ is the gravitational constant, and $R_{p}= 0.577\,\mathrm{pMpc}$ and $\sigma_{p}= 361\,\mathrm{km\, s^{-1}}$ are the projected radius and LOS velocity dispersion respectively, as defined in \citet{girardi:1998}).
This yielded a
core virial mass of $M_V= 8.2 \times 10^{13}\,M_\odot$.  
This estimate assumed that the protocluster core is already virialized,  {which would set an upper limit on the amount of hot intra-group gas that might be present --- if these galaxies do not form a virialized halo, the amount of extended hot gas would be significantly less}. We also cannot discount the possibility of the velocity spread being caused by the galaxies being lined up in a filament along the LOS over several Mpc. 
However, since the COSMOS-BIRTH density reconstruction technique takes peculiar velocities into account, both possibilities (lack of virialization or a LOS filament) are in principle included in the posterior results. In other words, we are confident that COSTCO-I will collapse into a cluster regardless of whether our estimate of the core properties is accurate.

\begin{figure*}[t!]
    \centering
    \includegraphics[width=7in,clip=true,trim=40 50 40 40]{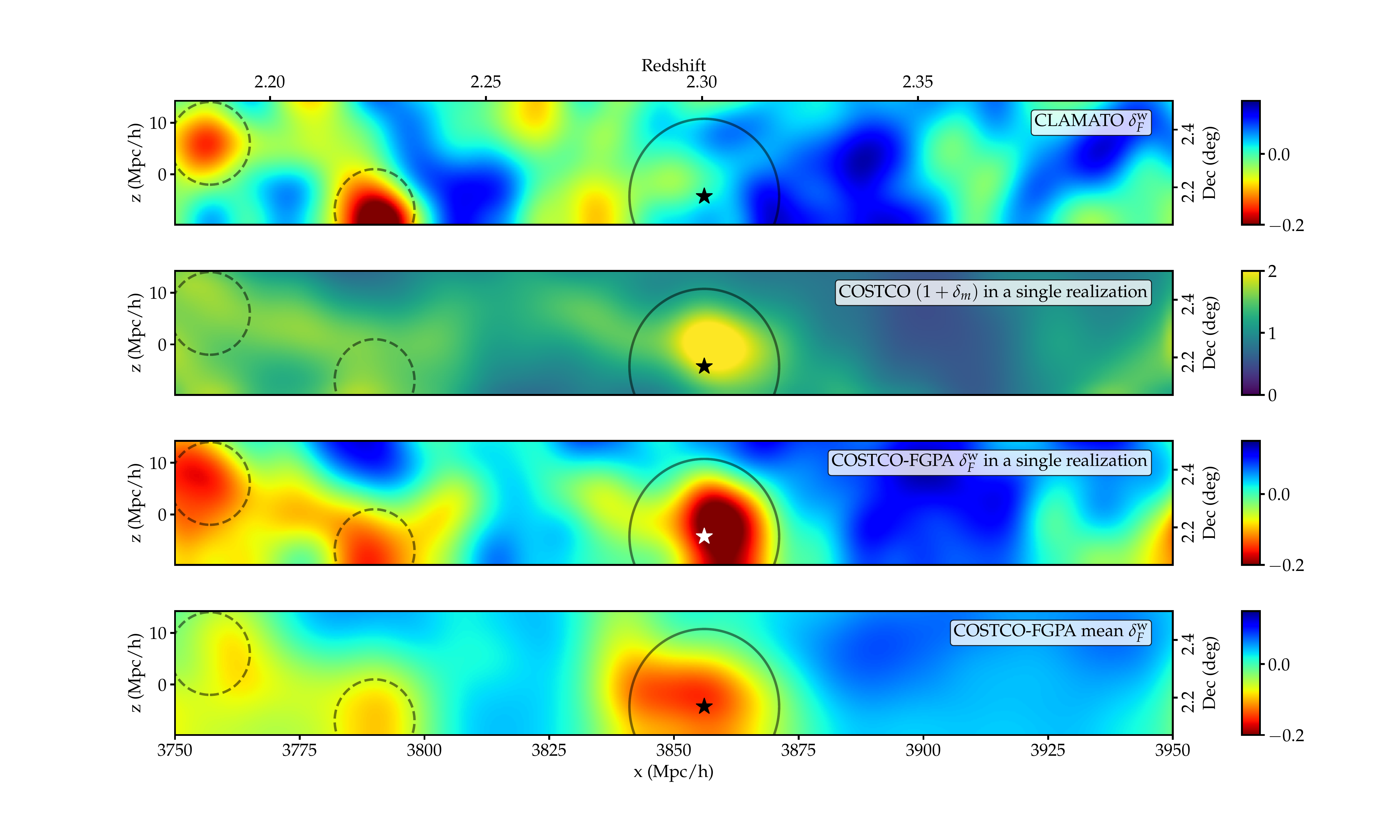}
    \caption{
        A series of smoothed sliceplots with thickness $2\,\hmpc$ ($150.095<$RA$<150.125$) that include the COSTCO-I galaxy protocluster, which is marked with the star. 
        From top to bottom, the panels show the transmission of CLAMATO, 
        the density contrast of one COSTCO constrained simulation, 
        the transmission of the corresponding COSTCO-FGPA Ly$\alpha$ forest mock and finally
        the mean Ly$\alpha$ transmission averaged over all realizations. 
        The abscissa (`$x$'-axis) is along the LOS dimension, while the ordinate (`$z$'-axis) is along increasing declination in the transverse plane.
        All the maps are smoothed with a $4\,\hmpc$ Gaussian kernel, while the circles indicate the $r=15\,\hmpc$ sphere over which we average in Section \ref{ssec:detection}. 
        In addition, we mark with dashed circles two other extended structures around the COSTCO-III (at $z=2.18$) and CC2.2 (at $z=2.22$) protoclusters, even though their barycenters are outside the plane of the slices shown here.
        %Note the strong absorption signal at $x\approx 3790\,\hmpc$ in the CLAMATO map, which corresponds to a different protocluster (CC2.2 at $z=2.232$) reported in \citet{ata:2022} --- although the protocluster center is beyond the boundaries of the CLAMATO map.
        }
    \label{sliceplot}
\end{figure*}

For this analysis, we also had in hand the $z=2.30$ matter overdensity field, $\delta_m = \rho/\bar{\rho} -1 $, where $\rho$ is the matter density, from 57 constrained realizations of the COSTCO $N$-body simulation suite that was designed to match the observed COSMOS galaxy distribution. 
These simulation outputs have a box size of $L_\mathrm{box}=512\,\hmpc$ and are binned in $256^3$ grid cells, covering the COSMOS volume in the redshift range $2.00<z<2.52$. The second panel of Figure~\ref{sliceplot}
shows the matter density contrast from one realization of COSTCO in the vicinity of the COSTCO-I protocluster.
Note that the barycenter of COSTCO-I in this particular COSTCO realization shown in Figure~\ref{sliceplot} is slightly offset from the reported position by \citet{ata:2022}, which comes from averaging over all the realizations in the COSTCO suite. 
This illustrates the fact that the ensemble of COSTCO realizations represents a posterior sample encapsulating our uncertainties regarding the protocluster masses and positions.

\subsection{Ly$\alpha$ Forest Absorption Data}

We now briefly describe the Ly$\alpha$ forest absorption maps that we used to study
the large-scale gas in the COSTCO-I protocluster, which is from the CLAMATO 
survey \citep{lee:2014a,lee:2018,horowitz:2022}. 

The CLAMATO survey was a spectroscopic survey that
targeted $z \sim 2-3$ UV-bright background sources that probe the Ly$\alpha$ forest in the COSMOS field,
using the LRIS spectrograph on the Keck-I telescope.
For the first time, star-forming galaxies were also systematically targeted as background sources in addition to the traditional quasars.
This enabled a high density of Ly$\alpha$ forest sightlines on the sky (857 deg$^{-2}$) which yielded a mean transverse separation of 2.35$\,\hmpc$ over a footprint of $\sim$0.2 deg$^2$ in the center of the COSMOS field.

The raw spectra were reduced, and then the unabsorbed continuum $C$ was estimated using the mean-flux regulation technique 
\citep{lee:2012}.
After selecting spectra with a continuum-to-noise ratio (CNR) 
criteria $\langle \mathrm{CNR} \rangle \ge 1.2$, 
the final sample for 3D reconstruction comprised of 320 galaxies and quasars in total.
For each spectral pixel in the rest frame $1041 \AA < \lambda < 1185 \AA$, the Ly$\alpha$ transmitted flux is defined as 
\begin{equation}
    \delta_F = \frac{f}{C \langle F \rangle (z)} - 1,
\end{equation}
where $\langle F \rangle (z)$ is the average transmission at a given redshift $z$.
The processed pixel data was mapped to a comoving volume covering 
$34\,\hmpc \times 28\,\hmpc \times 438 \,\hmpc$ along the R.A., Dec, and line-of-sight dimensions, respectively.

The Wiener filtering algorithm \verb|dachshund| \citep{stark:2015} was then applied on
the pixel data to create a reconstructed map of the 3D Ly$\alpha$ absorption $\delw$ in the redshift range $2.05<z<2.55$. 
The correlation lengths adopted for this reconstruction are $L_\perp =2.5\,\hmpc$ and $L_\parallel =2.0\,\hmpc$ in the transverse and line-of-sight dimensions, respectively. 

Note that as part of their CLAMATO data release, \citet{horowitz:2022} also reconstructed the underlying 3D matter density field using a different constrained realization method \citep{horowitz:2021a} from COSTCO. 
However, this used a combination of coeval galaxy positions in addition to the Ly$\alpha$ absorption as its input, assuming an FGPA-type relationship. 
Since we would like to use only the Ly$\alpha$ forest transmission for this analysis, we will use the Wiener-filtered Ly$\alpha$ map instead of the matter density map.
The 3D contours in Figure~\ref{plot3d} indicate the Ly$\alpha$ absorption in the vicinity of COSTCO-I.
It is clear that the region in the vicinity of COSTCO-I exhibits only 
average absorption ($\delw \sim 0$) instead of the strong absorption expected of a protocluster
\citep{stark:2015, qezlou:2022}

\begin{comment}
\subsection{Reconstruction of $\delta_F$ field} \label{ssec:recon}
%% WF / dachshund
We reconstruct $\delta_F^{rec}$ in the whole CLAMATO volume with from mock skewers 
via public available code \verb|dachshund| \citep{RN138}, which implement Wiener Filter reconstruction algorithm:
\begin{equation}\label{WF}
    \delta^{rec}_F = \mathbf{C}_{MD} (\mathbf{C}_{DD} + \mathbf{N})^{-1} \delta_F
\end{equation}
where $\delta^{rec}_F$ is the reconstruction of $\delta_F$ field
$\mathbf{C}_{MD}$ is the map-data covariances, 
$\mathbf{C}_{DD} + \mathbf{N}$ is the data-data covariance.

The implementation of \verb|dachshund| assumes 
the noise covariance matrix is diagonal
$N_{ii} = \sigma_i^2$, while 
$\mathbf{C}_{DD} = \mathbf{C}_{MD} = \mathbf{C}(\mathbf{r_1}, \mathbf{r_2})$, in which
\begin{equation}\label{noise_matrix}
    \mathbf{C}(\mathbf{r_1}, \mathbf{r_2}) = \sigma_F^2 \exp{[-\frac{(\Delta r_{\parallel}^2)}{2L^2_{\parallel}}]} \exp{[-\frac{(\Delta r_{\perp}^2)}{2L^2_{\perp}}]}
\end{equation}
With the formulation (\ref{WF}) and (\ref{noise_matrix}), 
\verb|dachshund| can solve the transmission map in the whole CLAMATO volume for large scale structures study.
The final product of \verb|dachshund| is a 3D-array representing the $\delta_F$ value in the CLAMATO volume.
The $\delta_F$ output carries information on the distribution of HI content in the volume; 
in the case that preheating is absent, 
the $\delta_F$ also has an indirect connection with underlying matter overdensity via (\ref{FGPA}).
\end{comment}

\section{Analysis} \label{sec:analysis}

The top panel in Figure~\ref{sliceplot} shows a narrow projected slice of the reconstructed Ly$\alpha$ transmission from CLAMATO, centered on the COSTCO-I coordinates, in comparison with a single realization of the matter density field that was constrained from galaxy tracers by COSTCO (second panel). 
In this figure, not only do we not see significant Ly$\alpha$ absorption associated with COSTCO-I, we see significant absorption features 
associated with two other protoclusters reported by \citet{ata:2022}: at $z=2.18$ we see the extended signal from COSTCO-III, 
while at $z=2.22$ there is a signal associated with the CC2.2 protocluster (\citealt{darvish:2020}, but also detected in COSTCO). 
In both cases, the protocluster centers actually lie outside of the map region shown in the figure, which further emphasizes the lack of signal associated with COSTCO-I. 

This lack of Ly$\alpha$ absorption associated with a known galaxy protocluster appears to depart from the FGPA (Equation~\ref{FGPA}), 
in which matter overdensities are expected to yield strong Ly$\alpha$ absorption  {on scales of $\sim$Mpc or greater}.
We also checked the preliminary maps
from the LATIS Survey \citep{newman:2020}, 
 a completely independent IGM tomography survey which also targeted the 
COSMOS field. A visual inspection of their Figure~26 shows no excess Ly$\alpha$ absorption in the vicinity of COSTCO-I.

We now proceed to quantify this discovery by adopting as our null hypothesis that
the protocluster gas associated with COSTCO-I follows the FGPA. 
The COSTCO constrained simulations offers a convenient way to test this null hypothesis: 
since its matter density field was estimated using galaxies as tracers, we can `paint' the Ly$\alpha$ absorption using FGPA, 
and incorporate the observational uncertainties of CLAMATO (e.g.\ sightline sampling, pixel noise, Wiener filtering). 

\subsection{COSTCO-FGPA Mock IGM Maps}

We use 57 COSTCO realizations at the $z = 2.3$ snapshot to generate the mock IGM tomography data that is matched to CLAMATO observational properties.
To convert real-space matter density of COSTCO 
into redshift-space transmission $F_{sim}$,
we make use of the FGPA relation (Equation~\ref{FGPA}),
In our calculation, we adopt the widely-used value $\beta = 1.6$ (e.g., \citealt{kooistra:2022a}).
Note the $\tau$ value here is extracted from real-space; 
to be consistent with the observations, we shift the $\tau$ to the redshift-space value $\tau_{red}$, 
and then compute the Ly$\alpha$ transmission in redshift space from 
\begin{equation}
    F_{sim} = \exp{(-\tau_{red})}
\end{equation}
As the proportional coefficient of the FGPA relation (Equation~\ref{FGPA}) is yet to be determined, we keep it as an unknown 
and solve its value by setting a mean transmission value \citep{becker:2013} 
\begin{equation}
    \langle F_{sim} \rangle = \langle F \rangle_{z=2.30} = 0.8447
\end{equation}
%For the sake of fast processing, we clip the central $100\,\hmpc \times 100 \,\hmpc \times 512\,\hmpc$ from each COSTCO snapshot
%and compute the transmission $F_{sim}$ via FGPA. 

Note that the resolution of the simulations is relatively coarse, with a 
grid of $2\,\hmpc$.
Therefore, the resulting FGPA transmission sightlines can not be expected to 
accurately reproduce the small-scale statistics of the Ly$\alpha$ forest. 
However, the Appendix of \citet{horowitz:2021a}
shows that even such a coarse grid resolution should suffice
to recover the structures of the cosmic web on scales of $\sim 2\,\hmpc$. We therefore do not expect the low resolution of the
COSTCO suite to significantly affect our analysis.

With the simulated FGPA transmission field $F_{sim}$ from the COSTCO suite in hand, 
we proceeded to generate mock skewers that reproduce the observational properties of CLAMATO as closely as possible, through the following steps:

\begin{enumerate}
    \item We extract noiseless 1-dimensional $F_{sim}$ sightlines at the $[x,y,z]$ positions probed by the CLAMATO sightlines. This process incorporates the  positions of CLAMATO sightlines in the plane of sky, as well the finite lengths of the segments along each LOS based on the redshift of the background sources. We also applied the pixel masks that were used to mask metal-line absorption in the CLAMATO spectra.
    \item 
    Continuum errors were introduced using the process described by \citet{krolewski:2017}. This assumed that the continuum estimation process
    results in $10\%$ fluctuations in the observed transmission, i.e.\
    \begin{equation}
        F_{obs} = \frac{F_{sim}}{1 + \delta_{cont}},
    \end{equation}
    where $\delta_{cont}$ is a Gaussian random deviate with mean value $0$ and standard deviation $0.1.$
    \item 
    Random pixel noise was added based on the signal-to-noise ratio, $\mathrm{SNR}$ estimated for each individual CLAMATO sightline. 
    This resulted in a final transmitted flux
    \begin{equation}
        F = F_{obs} + N_{obs},
    \end{equation}
    where Gaussian random deviate $N_{obs}$ (with standard deviation $\sigma = F / \mathrm{SNR}$) is the noise term.
    Finally, we obtained $\delta_F = F / \langle F \rangle -1 $ and computed the noise $\sigma_F$ from the SNR value of each sightline.
    %% todo: add formulas for this part
\end{enumerate}

For each COSTCO realization, we repeated the final noise-addition step 20 times with different noise seeds, to enhance the size of our mock sample.
As a result, we have $57 \times 20 = 1140$ sets of mock skewers with identical spatial sampling and signal-to-noise properties as CLAMATO, that were all designed to be consistent with the \emph{galaxy} density field observed in the COSMOS field. We dub this the COSTCO-FGPA sample.
We compile the pixel positions, $\delta_F$ and $\sigma_F$, and fed them into \verb|dachshund| as was done with the real CLAMATO data.
In the Wiener reconstruction, we kept the correlation lengths, $L_{\parallel} = 2\,\hmpc$ and $L_{\perp} = 2.5\,\hmpc$, the same as that of CLAMATO.
In the Appendix, we compare the overall properties of COSTCO-FGPA with the real CLAMATO data.

The output of these mock reconstructions, $\delw$, thus constitutes our null hypothesis: based on our knowledge of the COSTCO-I protocluster from the observed galaxy distribution, 
the COSTCO-FGPA mock data represents what we expect to see if the associated Ly$\alpha$ forest follows FGPA. 
Moreover, the ensemble of 1140 mock realizations represents all our uncertainties regarding the protocluster properties estimated by COSTCO (mass distribution and position) as well as those stemming from CLAMATO (sightline sampling and pixel noise).

\subsection{Detection of Large-Scale Heating in COSTCO-I} \label{ssec:detection}
In the third and fourth panels of Figure~\ref{sliceplot} we show the COSTCO-FGPA maps for one realization and the ensemble mean, respectively. 
One can immediately see the difference of $\delw$ between the CLAMATO and COSTCO-FGPA.
The CLAMATO IGM transmission value at the position of COSTCO-I is close to the mean ($\delw \sim 0$), while from the COSTCO-FGPA realization one sees strong absorption ($\delw < -0.2$) at the same position.
The presence of absorption feature in the averaged $\delta_F$ map for COSTCO-FGPA further confirms the mock absorption feature.

\begin{figure*}[t!]
    \centering
    \includegraphics[width=0.75\textwidth,clip=true,trim=30 90 30 110]{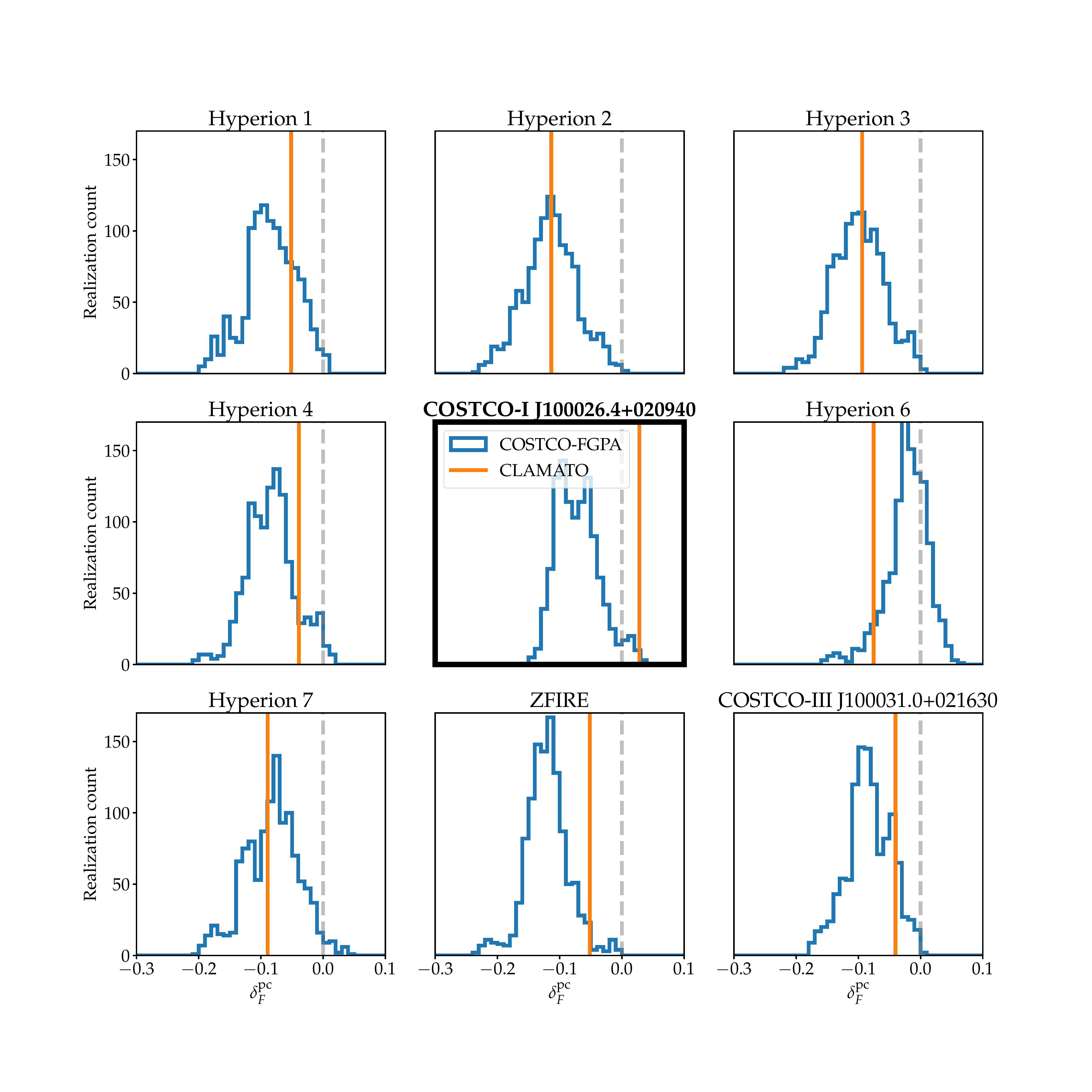}
    \caption{
        The distribution of $\delpc$ derived from the method in Section \ref{ssec:detection}.
        We include all the 9 protoclusters in the overlap of CLAMATO and COSTCO volumes. 
        The blue histogram is the distribution of $\delpc$ in 1140 COSTCO-FGPA mock realizations;
        the orange lines mark the $\delpc$ of corresponding protoclusters in the observed CLAMATO;
        the grey vertical dash lines represent $\delpc = 0$.
        We highlight COSTCO-I with bold text (the center plot) --- the mocks for this structure are clearly inconsistent with the CLAMATO measurement.
    }
    \label{Detection_preheat}
\end{figure*}

As the COSTCO-FGPA $\delw$ shares the same shape and coordinate system with CLAMATO $\delw$,
we can perform a direct, quantitative comparison between the CLAMATO and any COSTCO-FGPA realization at the position of the protocluster.
First, we smooth the $\delw$ maps with a $R = 4\,\hmpc$ top-hat kernel \citep{stark:2015a}, 
and then compute the mean $\delw$ value enclosed by a $R=15\,\hmpc$ sphere centered at the reported COSTCO-I position; we refer to this quantity as $\delpc$. 
The radius is inspired by \citet{ata:2022}, who defined a protocluster as structure that consistently formed a $z=0$
cluster within a $R = 15\,\hmpc$ radius of each other in the $z=0$ snapshots.
% The average may reduce the significance of detection 
% because this radius is beyond the typical size of protoclusters at $z \sim 2$.

We computed $\delpc$ centered on the protocluster for all 1140 COSTCO-FGPA realizations, and compared them with the value computed from CLAMATO.
This distribution is shown in the central panel of Figure~\ref{Detection_preheat}.
The CLAMATO Ly$\alpha$ transmission associated with COSTCO-I is clearly 
more transparent (less absorbed) than seen in the COSTCO-FGPA mocks: we find that only 3 out of a total of 1140 COSTCO-FGPA realizations exhibit a $\delpc$ value
greater than seen in CLAMATO. The COSTCO-FGPA mocks represent the null hypothesis that the gas in COSTCO-I is following FGPA, for which we find a probability of $p = 1-3/1140=0.00263$ (corresponding to $2.79\sigma$ assuming a Gaussian distribution)
after incorporating all known uncertainties.
This is well below the standard hypothesis testing threshold of $p=0.05$, indicating a clear rejection of the null hypothesis: the protocluster gas in COSTCO-I does not follow the FGPA. 
 {Based on Equation~\ref{FGPA}, the reduced Ly$\alpha$ optical depth (i.e.\ increased transmission) in COSTCO-I might be due to an increase in the large-scale
gas temperature, or enhanced local UV background. }

COSTCO-I is not the only COSTCO-detected protocluster that falls within the CLAMATO volume.
We therefore also computed $\delpc$ for these other protoclusters, 
which include well-known structures such as the ZFIRE protocluster at $z=2.11$ \citep{nanayakkara:2016} and the various peaks of Hyperion at $z\approx 2.45 - 2.52$ \citep{cucciati:2018}.
These are shown in the non-central panels of Figure~\ref{Detection_preheat}, 
where the CLAMATO Ly$\alpha$ transmission is generally consistent with the COSTCO-CLAMATO mocks. 
This suggests that a transparent Ly$\alpha$ forest in 
protocluster gas is not an ubiquitous process at this epoch, 
although with we cannot rule out more subtle deviations from FGPA with the current data.
Indeed, there are hints that the ZFIRE protocluster might not obey FGPA (bottom-central panel of Figure~\ref{Detection_preheat}), but more detailed studies would be needed to confirm this.

The transparent Ly$\alpha$ forest of COSTCO-I extends across physical scales of $> 4\,$pMpc. We have identified a compact overdensity of galaxies that forms the putative protocluster core, which would have a characteristic radius of $r_{200}= 400\,\mathrm{pkpc} = 0.92\,\hmpc$ based on our estimated virial mass of $M_V = 5.78 \times 10^{13}\,\hmsun$. This is considerably smaller than the extent of the transparent gas
we see in the protocluster and is therefore unlikely to be due to early formation of an intra-cluster medium (ICM). 
Indeed, previous studies of hot gas possibly associated with ICM formation at $z\sim 2$ \citep[e.g.,][]{wang:2016,champagne:2021} were on much smaller scales of $\sim 100$kpc.
In any case, \citet{lee:2016} had tested a toy model in which the Ly$\alpha$ transmission was set to 100\% transmission ($F\equiv\exp(-\tau)=1$) within a
$1.5\,\hmpc$ radius of a simulated protocluster but with the gas outside following FGPA. 
This had a negligible effect on the averaged $\delpc$ computed over several Mpc, so a virialized ICM with $r_{200}=0.92\,\hmpc$ cannot be responsible for the spatially extended deviation from FGPA.

We believe there are three possibilities for this large-scale ($>4$pMpc) protocluster heating.
The first scenario is that the protocluster gas is 
being collisionally shock-heated due to gravitational collapse of the accreting material on large scales. 
However, this seems disfavored by prior theoretical analysis of the large-scale Ly$\alpha$ forest signal in $z\sim2$ protoclusters. \citet{miller:2021} studied $z\sim2.4$ protoclusters within the IllustrisTNG100 hydrodynamical simulation \citep{weinberger:2017,pillepich:2018}. 
They found that for an uniform UV background, the effect of collisional ionization is negligible on the smoothed 3D Ly$\alpha$ forest
signal associated with protoclusters. 
\citet{kooistra:2022}, on the other hand, performed zoom-in hydrodynamical simulations on a set of galaxy protoclusters, with various phenomenological 
 pre-heating prescriptions of the protocluster gas. They did however perform fiducial runs where gas hydrodynamics was in effect, but no feedback
 or pre-heating was applied. Even in $z=2$ protoclusters associated with the most massive $z=0$ clusters ($M(z=0) \sim 10^{15}\,\hmsun$), 
 simple gravitational shock heating appears to be generally insufficient to make the $\sim$Mpc-scale Ly$\alpha$ absorption significantly more transparent than the canonical FGPA. 
 These two studies, however, analyzed only a small number of simulated protoclusters: $\sim 20$ by \citealt{miller:2021} and 5 by \citealt{kooistra:2022a}. 
 Therefore, while they did not find large-scale gravitational 
 shock heating at $z\sim 2.3$, we cannot rule out the possibility that this is in progress in a small fraction of protoclusters. %, especially if additional protocluster galaxies in COSTCO-I might be UV-faint as recently discovered by \citet{newman:2022}. Such galaxies would not have been captured in the spectroscopic surveys used as the basis for the COSTCO study, leading to an underestimate of the full extent of the COSTCO-I protocluster. 
 A more comprehensive study involving large numbers of simulated protoclusters would help clarify this.

The second possibility for the large-scale heating is that feedback processes from protocluster galaxies or AGN is responsible for the reduced absorption in COSTCO-I. \citet{kooistra:2022} applied a simple phenomenological pre-heating model to protocluster gas, which imposed an entropy floor, $\kfloor$, such that protocluster gas cells at $z=3$ have internal gas entropy
values of $T \; n_e^{-2/3} > \kfloor $, where $T$ is the gas temperature and 
and $n_e$ is the electron density (see also \citealt{borgani:2009}). 
Their results showed that a significant entropy floor of $\kfloor \gtrsim 50$ keV cm$^{-3}$ would be required to cause increased Ly$\alpha$ transmission to $\delta_F \sim 0$ in $z\sim 2$ protoclusters. 
\citet{kooistra:2022} were agnostic on the possible mechanisms that could cause such preheating. 
Over the years, however, there has been a growing consensus that feedback from AGN 
is necessary to reproduce various properties related to galaxy clusters and groups \cite[e.g.,][]{puchwein:2008,mccarthy:2010}. 
AGN feedback is included in the IllustrisTNG100 simulation analyzed by \citet{miller:2021}, who showed that it does not appear to cause strong deviations from a FGPA-like relationship between the Ly$\alpha$ transmission and matter
density in $z=2.44$ protoclusters. 
However, the AGN feedback energy in the TNG model is deposited isotropically in the immediate vicinity of each SMBH. 
The AGN feedback in the Simba simulations \citep{dave:2019}, on the other hand, 
implements a collimated-jet feedback scheme for low-Eddington ratio AGN. This allows 
feedback energy to reach much larger scales compared with TNG \citep{tillman:2022}.
AGN jet feedback also appears to be driving large-scale heating at $z\sim 2$ in the HorizonAGN suite \citep{chabanier:2020} with significant effects on global Ly$\alpha$ forest statistics,
although they did not focus on protoclusters.
At low-$z$, radio lobes from giant radio galaxies have been shown to extend up to $\sim 4-5\,$Mpc \citep{delhaize:2021,oei:2022}, so similar mechanisms operating in high-redshift protoclusters could be heating up the proto-ICM over similar scales.

 {
Finally, an enhanced local UV radiation field, $\Gamma_\mathrm{uv}$,
from AGN within the protocluster could be an alternative cause of deviations from FGPA. However, \citet{miller:2021} also considered a radiative model with a realistic quasar luminosity function in their analysis of IllustrisTNG100 protoclusters, and showed that this is unlikely to denude the Ly$\alpha$ forest absorption of protoclusters to the extent that we see in COSTCO-I.
Rare hyper-luminous quasars (with absolute magnitudes of $M_{1450} \gtrsim -27$) would have a more significant effect on the Ly$\alpha$ absorption \citep{visbal:2008,schmidt:2018} than the simulated AGN considered by \citet{miller:2021}, 
but we see no Type-I quasars within this protocluster. 
On the other hand, obscured Type-II quasars emitting anisotropically away from our line-of-sight might indeed cause deviations from FGPA. 
 Hyper-luminous quasars are however rare in the Universe and we therefore deem it less unlikely cause of the Ly$\alpha$ transparency in COSTCO-I than collisional heating or jet-feedback.
It would however be an exciting discovery if an obscured hyper-luminous quasar were responsible for the transparency of COSTCO-I. We will investigate this possibility in a follow-up study of the protocluster members. 
}

\section{Conclusion} \label{sec:discussion}
% With two subsections:
% 1. estimation of PC halo mass
% 2. future ...

In this Letter, we presented evidence that the Ly$\alpha$ forest associated with an observed galaxy protocluster at $z=2.298$ is considerably more transparent (i.e.\ less absorbed) than expected given the overdensity of the protocluster.
We interpret this to be caused by elevated gas temperatures that
departs from the usual FGPA relationship that governs the Ly$\alpha$ transmission as a function of underlying matter density field. 

%In [22]: from colossus.halo import mass_so
%In [23]: mass_so.M_to_R(5.78e13, 2.3, '200m')
%Out[23]: 281.20024764902956

Whichever the true heating mechanism might be, the COSTCO-I galaxy protocluster appears to be the first known large-scale structure where the gas is undergoing the transition from the optically-thin photoionized temperature-density relationship of Cosmic Noon ($z\sim 2-4$), into the ICM by $z=0$.
In follow-up studies, we will study the effects of various feedback mechanisms in hydrodynamical simulations specifically in context of $z\sim 2-3$ protoclusters, while
also examining the multi-wavelength data extant in COSMOS field to search for trends in 
the constituent galaxies and gas associated with COSTCO-I.

\begin{comment}
\section{To-do}
\begin{itemize}
    \item Add transverse plot showing galaxy positions and CLAMATO $\delta_F$ (maybe an interactive one)\\
    Changelog: 2022.12.06 - Add a static figure for pdf version
    \item Add calculated numbers for protocluster core (with original method instead of NMAD; we change the choice of pc members, so need to add some text to illustrate; considering put the description \& galaxy catalog in the appendix)
    \item Recalculate $\delpc$ with active search for COSTCO density maximum in each realization (it makes the background of the plots hard to explain, so we will stick on the previous version)
    \item Reduce range of y-axis in $\delta_f$ PDF plot (fig~\ref{global hist}) (Now from $10^{-3}$ to $10^1$, fin)
    \item Clear up notation for the various different $\delta_F$\\
    For notations: 
    $\delta_F$ -> raw delta of skewers; also used as general notation in the discussion
    $\delta_F^w$ -> Wiener-Filtered stuff, both for CLAMATO and COSTCO
    $\delta_F^{\mathrm{pc}}$ -> Average value in spheres
    $\delta_m$ -> mass overdensity (use little m instead of M)
    \item Ensure the result (fractions of realizations) is consistent with current 15 Mpc/h average (fin)
    \item Use different notations of redshift to avoid confusion
    
    \item Add secondary [Dec, redshift] axes for sliceplots (fin)
    \item relabel $\delta_f$ as $\delpc$ in Fig~\ref{Detection_preheat} (switch to 15 Mpc/h average, fin)
\end{itemize}
\end{comment}
 
\section{Acknowledgments}
We thank Renyue Cen for useful discussions that helped initiate this project, and Mike Rich for useful feedback on the draft. 
Kavli IPMU was established by World Premier International Research Center Initiative (WPI), MEXT, Japan. C.Z.D. is supported by Forefront Physics and Mathematics Program to Drive Transformation (FoPM), a World-leading Innovative Graduate Study (WINGS) Program, the University of Tokyo. K.G.L. acknowledges support from JSPS Kakenhi grants JP18H05868 and JP19K14755. M.A. was supported by JSPS Kakenhi Grant JP21K13911. The data presented herein were obtained at the W.M.\ Keck Observatory, which is operated as a scientific partnership among the California Institute of Technology, the University of California and the National Aeronautics \& Space Administration (NASA). The Observatory was made possible by the generous financial support of the W.M. Keck Foundation. We also wish to recognize and acknowledge the very significant cultural role and reverence that the summit of Maunakea has always had within the indigenous Hawai’ian community. We are most fortunate to have the opportunity to conduct observations from this mountain.

\begin{figure}[h]
    \centering
    \includegraphics[width=0.5\textwidth,clip=true,trim=10 0 10 40]{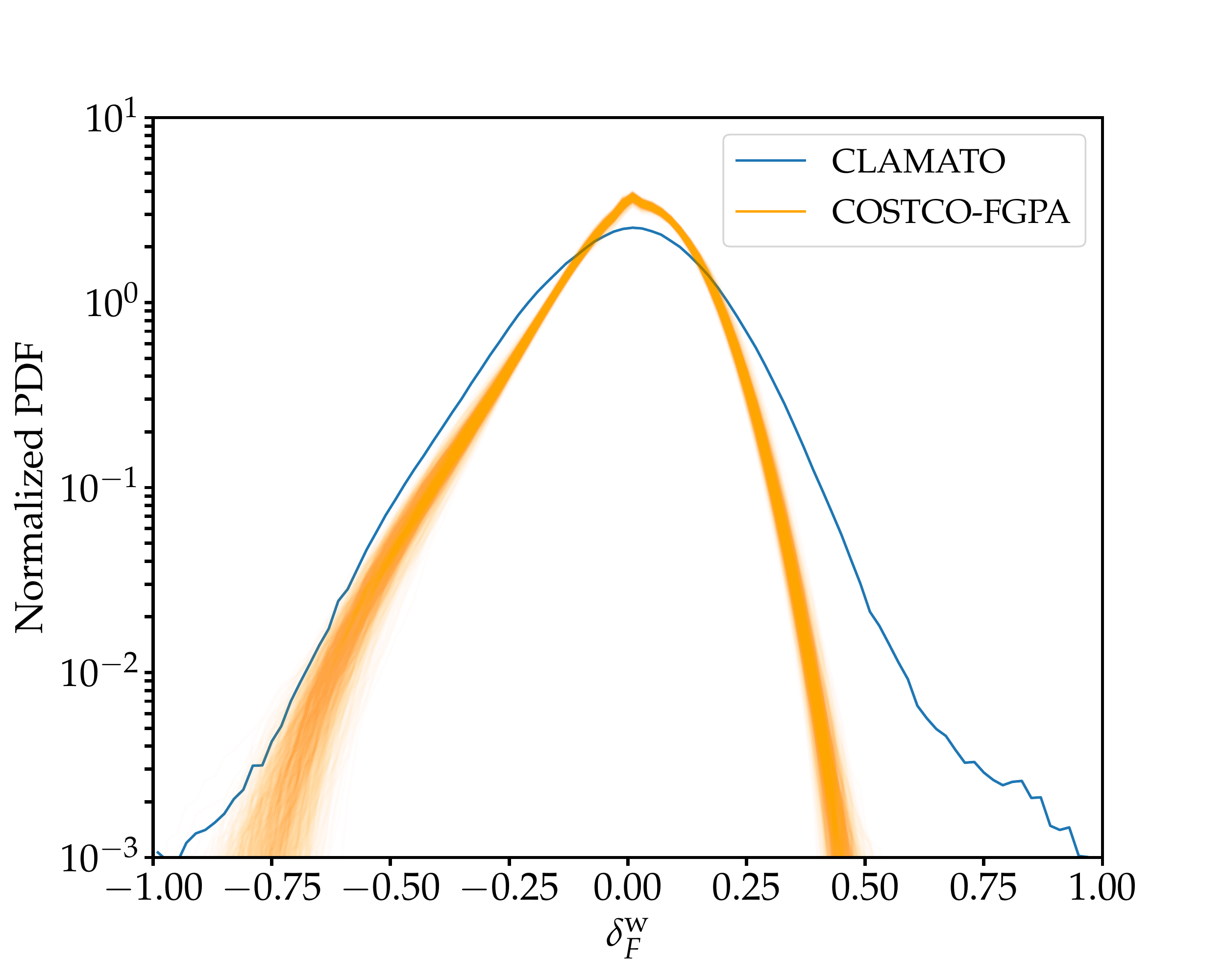}
    \caption{
        The distribution of $\delw$ values from Wiener-filtered Ly$\alpha$ absorption maps in the CLAMATO field. 
        The orange histograms are the distribution of different realizations from the COSTCO-FGPA reconstructions,
        while the blue histogram is for the actual CLAMATO data. The distributions agree reasonably well in the low-transmission regime ($\delw \lesssim 0$) which traces overdensities. 
        }
    \label{global hist}
\end{figure}

\appendix
In this Appendix, we compare the global properties of the COSTCO-FGPA
mock maps with the CLAMATO observational data.
In Figure~\ref{global hist}, we present the probability distribution function (PDF) of CLAMATO and COSTCO-FGPA transmission field.
We find that the CLAMATO and COSTCO-FGPA transmission distributions are in good agreement in the low transmission region, 
while the high transmission region is not well-reproduced in the mock transmission map.
We attributed this to the tracers used in COSTCO, which are galaxies with known spectroscopic redshifts.
According to the galaxy formation and evolution theory, 
galaxies are formed in the dark matter halos which are located in the density peaks.
This means that reconstructions at their most reliable  in high density (low Ly$\alpha$ forest transmission) regions;
on contrary, the structures in the low density regions are mostly introduced by noises of mock data.

\begin{figure*}[h]
    \centering
    \includegraphics[width=0.8\textwidth, clip=true,trim=50 40 30 50]{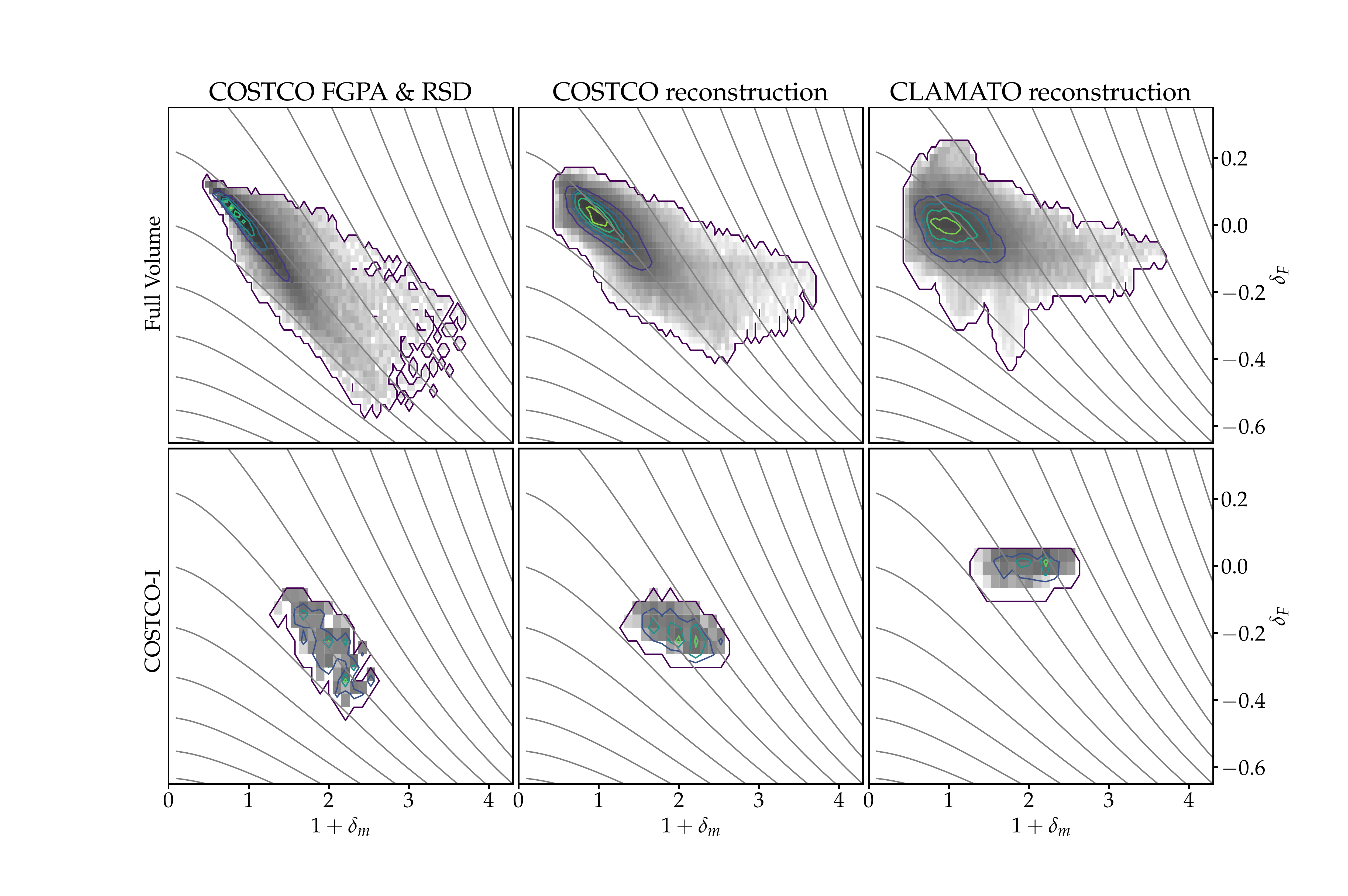}
    \caption{Top row: the transmission-density relation of the whole CLAMATO volume.
             Bottom row: the transmission-density relation of $4.5\,\hmpc$ sphere centered at COSTCO-I.
             Left-most column shows the relations of underlying matter density versus $\delta_F$ after applying FGPA \& redshift-space distortions to the COSTCO matter density field; 
             Middle column: the same after applying sightling sampling and Wiener reconstruction; Right column: the Wiener-reconstructed Ly$\alpha$ flux as seen in the real CLAMATO data. 
             The grey lines indicates the FGPA relation $\tau \propto (1 + \delta)^{1.6}$.
             All the data are spatially smoothed with a $4\,\hmpc$ Gaussian kernel.}
    \label{TDR_fig}
\end{figure*}

As proposed by \citet{kooistra:2022a}, the Ly$\alpha$ transmission-density relation can be a diagnostic of galaxy feedback.
Figure~\ref{TDR_fig} depicts this relation for the whole CLAMATO volume and the COSTCO-I region.
The left column shows the relationship from one COSTCO-FGPA realization without noise, 
finite sightline sampling, or Wiener-filter. We find that the FGPA relation is tightly preserved after applying redshift distortions (i.e.\ peculiar velocities) in both the full volume and near COSTCO-I;
After picking sightline sampling, adding noise, and Wiener-Filtering, the majority of the whole volume still follows the FGPA relation, 
while the pixels around COSTCO-I show a slightly different trend.
We regard it as a consequence of information loss during the generation of mock skewers.
The transmission and density in the full CLAMATO volume do show a correlation, 
but it is not consistent with FGPA relation of $\beta = 1.6$.
This is likely due to additional uncertainties in the matter density construction ($\delta_M$) that is not reproduced in the COSTCO-FGPA relationship. 
Around COSTCO-I, we do not find a notable trend between the transmission and density  -- the $\delta_F$ values remain almost constant. This is reminiscent of the flat transmission-density 
relations that have high levels of pre-heating as studied by \citet{kooistra:2022}. 
However, more quantitative studies involving the transmission-density relationship would require large observational samples and more careful modelling of systematics.

\bibliography{references_kg}{}
\bibliographystyle{aasjournal}



\end{document}